\documentclass[11pt]{article}
\usepackage{amsmath}
\usepackage{graphicx}
\usepackage{bm}

\begin{document}

\title{Isotopically engineered silicon nanostructures in quantum computation and
communication\footnote{Presented at International Workshop 
\textit{Frontiers in Science and Technology}. Holon Academic Institute of 
Technology, Holon, Israel, 26--27 October 2003}}
\author{Issai Shlimak \\
%EndAName
\\
{\small {\textit{Jack and Pearl Resnick Institute of Advanced Technology,}}}%
\\
{\small {\textit{Department of Physics, Bar-Ilan University, Ramat-Gan
52900, Israel}}}\\
{\small {\textit{e-mail: shlimai@mail.biu.ac.il}}}}
\date{}
\maketitle

\thispagestyle{empty}

\begin{abstract}
Natural silicon consists of three stable
isotopes with atomic mass 28 (92.21\%), 29 (4.70\%) and 30 (3.09\%). To
present day, isotopic enrichment of Si was used in electronics for two
goals: (i) fabrication of substrates with high level of doping and
homogeneous distribution of impurities and (ii) for fabrication of
substrates with enhanced heat conduction which allows further chips
miniaturization. For the first purpose, enrichment of Si with Si$^{30}$ is
used, because after irradiation of a Si ingot by the thermal neutron flux
in a nuclear reactor, this isotope transmutes into a phosphorus atom which
is a donor impurity in Si. Enrichment of Si with Si$^{30}$ allows
one to increase the level of doping up to a factor of 30
with a high homogeneity of the impurity distribution. The second purpose is
achieved in Si highly enriched with isotope Si$^{28}$,
because mono-isotopic Si is characterized by enhanced thermal conductivity.

New potential of isotopically engineered Si comes to light because of
novel areas of fundamental and applied scientific activity connected with
spintronics and a semiconductor-based nuclear spin quantum computer where
electron and/or nuclear spins are the object of manipulation. In this case,
control of the abundance of nuclear spins is extremely important.
Fortunately, Si allows such a control, because only
isotope Si$^{29}$ has a non-zero nuclear spin. Therefore, enrichment or
depletion of Si with isotope
Si$^{29}$ will lead to the creation of a material with
a controlled concentration of nuclear spins.
Two examples
of nano-devices for spintronics and quantum computation, based on isotopically engineered silicon,
are discussed.\newline
\end{abstract}

\newpage
%\bigskip

Development of semiconductor technology is determined by industry
requirements. The most exploitable material in the modern electronics and
semiconductor industry is Si. More than 90\% of semiconductor devices on
the market are made from silicon. Two aspects of industrial application
caused the development of the isotopic engineering of Si.

The first one is connected with the problem of homogeneous doping of Si.
The tendency in the microelectronic technology is directed to use the Si
slices with bigger diameter to prepare more chips in one process. As a
result, Si ingots with diameter of 200 and even 300~mm are
fabricated now by semiconductor industry. The Si slices with big diameter
are used also in fabrication of high power electrical $ac/dc$ converters. In
both applications, high homogeneity of doping of slices is of crucial
importance. Meanwhile, in conventional metallurgical methods of doping where
the impurity is introduced into the melt with subsequent growth of the
semiconductor crystal, obtaining a homogeneous distribution of an impurity
encounters radical difficulties. They are associated with the instabilities
in the frontline of crystallization of doped semiconductors and an
unavoidable temperature gradient in the growing ingot between its center and
periphery. These difficulties increase dramatically as the ingot diameter
increases. Therefore, increase of the ingot diameter leads to increase of
heterogeneity in the impurity distribution.

By contrast, neutron transmutation doping (NTD) of Si is characterized by
the very high homogeneity in impurity distribution. The NTD method is based
on nuclear transformation of isotopes of semiconductor materials following
their capture of slow (thermal) neutrons [1,2]. NTD is achieved by
irradiation of semiconductor crystal with a neutron flux in a nuclear
reactor. On capturing a neutron, a particular isotope transmutes to another
isotope with a mass number larger by one:
\begin{equation}
_{Z}N^{A}\sigma _{i}\Phi =_{Z}N^{A+1}  \label{eq1}
\end{equation}
Here $\Phi$~(cm$^{-2}$)
is the integrated flux (dose) of thermal neutrons,
$\sigma _{i}$~(cm$^{2}$) is the thermal-neutron-capture cross section for a
given isotope, $_{Z}N^{A}$ and $_{Z}N^{A+1}$~(cm$^{-3}$)
are concentrations
of initial and final reaction products, respectively, $Z$ is the nuclear
charge, and $A$ is the mass number of the nucleus. If the isotope thus
obtained, $_{Z}N^{A+1}$, is stable, such nuclear reaction does not entail
doping. Of most interest is the case, where the final isotope is unstable.
Then after its half-life, this isotope transmutes to a nucleus of another
element, with nuclear charge (atomic number) is larger by one,
$_{Z+1}N^{A+1} $ as in the case of $\beta ^{-}$ decay, or smaller by one,
$_{Z-1}N^{A+1}$, if the decay occurred by $K$-capture of the intrinsic
electron. One may present here for illustration the reaction producing in
silicon a phosphorous donor impurity:
\begin{equation}
_{14}{\rm Si}^{30}+_{0}n^{1}=_{14}{\rm Si}^{31}\rightarrow
\beta (2.62h)\rightarrow _{15}{\rm P}^{31}  \label{eq2}
\end{equation}
In the case of Si, which consists of three stable isotopes, Si$^{28}$
(abundance 92.2\%), Si$^{29}$ (4.7\%) and Si$^{30}$ (3.1\%), the NTD method
has two main advantages over the conventional metallurgical methods of
impurity incorporation. First, this is high-precision doping, because the
concentration of impurities introduced at a constant neutron flux is
proportional to irradiation time, which can be controlled with a high
accuracy. The second advantage is the high homogeneity of doping [3]. This
is provided by the random distribution of isotope Si$^{30}$ in the crystal
lattice, the uniformity of the neutron flux (to achieve this, the ingot is
rotated about its axis and pulled simultaneously through the reactor active
zone during the irradiation), and the small value of $\sigma _{i}$. As a
result, high macro-homogeneity of doping was achieved even for ingots with
diameter of 200~mm (Fig.~1).

\begin{figure}[t]
\begin{center}
	\includegraphics{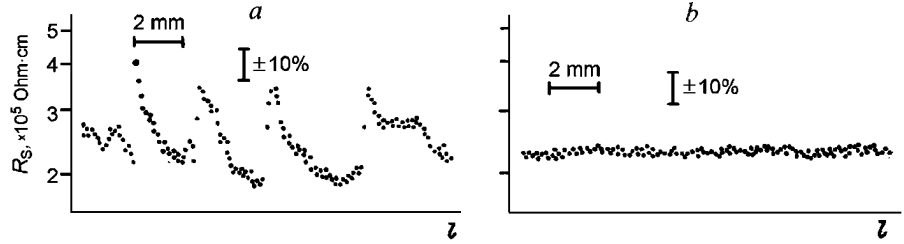}
	\caption{Microdistribution of resistivity in conventionally doped Si (a) and
			in NTD-Si (b) [3].}
\end{center}
\end{figure}

However, the NTD method has a limitation, connected with
existence of the upper (maximal) level of doping. Indeed, concentration of
the phosphorous impurity $N$[P$^{31}$] introduced into Si using NTD method
is determined by Eq. (1):
\begin{equation}
N[{\rm P}^{31}]=N[{\rm Si}]x^{30}\sigma _{i}\varphi t  \label{eq3}
\end{equation}
Here $N[{\rm Si}] = 5\times 10^{22}$~m$^{-3}$
is the density of Si atoms in a
lattice, $x^{30}$ = 0.031 is the abundance of isotope Si$^{30}$,
$\sigma_{i} = 0.11\times 10^{-24}$~cm$^2$
is the cross-section for thermal neutron
capture by isotope Si$^{30}$, $\varphi $ is the thermal neutron flux, $t$ is
the time of irradiation. In conventional nuclear reactors, where the value
of $\varphi $ do not exceed usually
$5\times 10^{13}{\rm~cm}^{-2}{\rm~s}^{-1}$, one month of
irradiation corresponds to $2.6\times 10^{6}$~s,
therefore maximal flux is about
$10^{20}$~cm$^{-2}$ and therefore the maximal level of doping is limited by the
value of $N[{\rm P}^{31}] \approx 10^{16}{\rm~cm}^{-3}$.
Longer time of irradiation
is unwanted because of accumulation of unremovable radiation damages. In
this case, the only way to achieve a higher level of doping is the isotopic
enrichment of Si with Si$^{30}$. Taking into account that the abundance of
Si$^{30}$ in natural Si is rather small (3.1\%), enrichment of Si with
Si$^{30}$ can increase the maximal level of NTD doping significantly, up to
the factor of 30.

The next field where isotopically engineered Si can be used in electronics
is a fabrication of Si substrates with enhanced thermal conductivity. This
is an important parameter in fabrication of integrated circuits (IC),
because it limits the density of transistors per unit area of substrate.
Thermal conductivity of Si is rather high which is an additional reason
why this material is widely used in electronics. However, in accordance with
the Technology Roadmap for Nanoelectronics [4], it is expected that in 2012
year, the industry will provide IC with over 10$^8$ transistors per
cm$^2$ for logic and over 10$^{10}$ bits per cm$^2$ for memory. Such
extremely high density of switching elements requires a substrate with
enhanced thermal conductivity to take away the emitted heat. In this case,
enrichment of Si with one isotope seems to be an effective way to increase
the thermal conductivity because variation of masses of different isotopes
leads to additional scattering of phonons. Experiments with mono-isotopic
Si, enriched with the main isotope Si$^{28}$ up to 99.7--99.896\% [5,6]
showed, indeed, an increase of thermal conductivity at room temperature up
to 60\% in comparison with natural Si (Fig.~2).

\begin{figure}[t]
\begin{center}
	\includegraphics{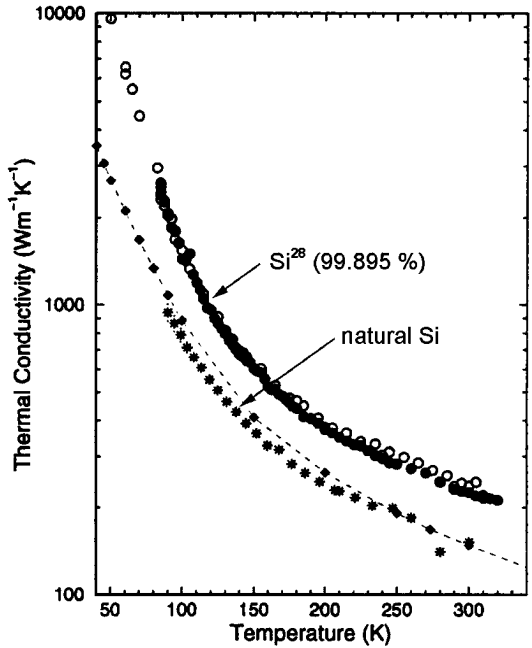}
	\caption{Thermal conductivity data for natural Si and
isotopically enriched Si$^{28}$ (99.895\%) [6].}
\end{center}
\end{figure}

New potential of isotopically
engineered Si come to light because of novel areas of fundamental and
applied scientific activity connected with spintronics and
semiconductor-based nuclear spin quantum computer where electron and/or
nuclear spins are the object of manipulation. In this case, control of the
abundance of nuclear spins is extremely important, because even if the
electron spin only is considered as a carrier of information, interaction
with nuclear spin system determines the electron spin decoherence time.
Fortunately, Si allows such a control, because only isotope Si$^{29}$ has
non-zero nuclear spin. Therefore, enrichment or depletion Si with isotope
Si$^{29}$ will lead to the creation of a material with a controlled
concentration of nuclear spins, and even without nuclear spins. One might
deliberately vary the isotopic composition to produce layers, wires and dots
that could serve as nuclear spin qubits with a controlled number of nuclear
spins.

Let us consider an example, where isotopic engineering of Si allows to
suggest a basic building block of quantum computation (QC) a technology
viable two-quantum bit (qubit) device for semiconductor-based nuclear spin
quantum computer (NSQC) [7]. Nuclear spins are the leading candidate for
storing and manipulating QC information because they are well isolated from
their environment. Therefore, operations on nuclear spin qubits could have
low error rates, and nuclear spin-based data storage elements can have long
retention times. The information encoded in the nuclear spin polarization
can be detected by an electrical measurement. An ensemble of nuclear spin
will be polarized which means that during this stage, we are concerned with
spintronics, rather than QC. However, the use of isotopically engineered Si
enables us to reduce the abundance ratio for nuclear spin isotopes in a
typical size of a quantum dot (10$\times$10~nm) to any small number.
The states
of the qubit will be encoded in the polarization of nuclear spins in a
system of two coupled quantum dots (QDs) fabricated from a Si/SiGe
heterostructure containing a high-mobility two-dimensional electron gas in
the quantum Hall effect regime. The nuclear spin polarization will be
accomplished via the hyperfine interaction with electron spin system. The
same electrons will nondissipatively transfer the polarization between QDs,
as well as permit the reading out of the polarization via a
magnetoresistance measurement [8--10].

\begin{figure}[t]
\begin{center}
	\includegraphics{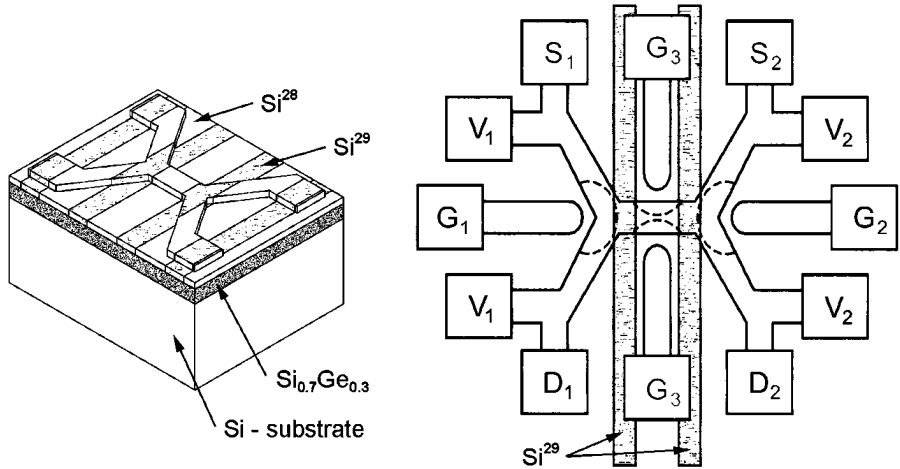}
	\caption{Two nuclear-spin-qubit device made from isotopically engineered
Si/SiGe heterostructure.}
\end{center}
\end{figure}

The key feature of the proposed structure is a stripy Si layer
which consists of a sequence of strips of non-zero nuclear spin isotope
Si$^{29}$ (or natural Si with 4.7\% of Si$^{29}$) and isotope Si$^{28}$
without nuclear spin (Fig.~3). Using appropriate negative voltage applied to
the gate electrodes $G_{1}$, $G_{2}$ and $G_{3}$, one can squeeze out
electrons from close-fitting areas of the structure and form a two-qubit
system, which consists of two QDs with a few nuclear spins and no nuclear
spins in space between dots.

Fabrication of the proposed devices does not present an insurmountable task.
Striped Si layer can be obtained using molecular beam epitaxy (MBE) growth
on vicinal substrates using multiple Si sources [11]. The only difficulty
is the proper alignment of the coupled quantum dot device with respect to
the isotopically engineered stripes in the Si quantum well. For this
purpose it is suggested to use the nanoscale patterning under the control of
atomic force microscope (AFM) [12]. This microscope will be used also for
visualization of the structure steps on the surface of grown heterostructure.

Isotopically enriched Si and Ge compositions are, in principle,
available. For example, the properties of highly enriched (99.896\%) 
Si$^{28}$ were reported in Ref. [6]. In this isotopically enriched Si, the
abundance of Si$^{29}$ and Si$^{30}$ together is as small as 0.104\%.
Therefore, the abundance of non-zero nuclear spin isotope Si$^{29}$ is
expected to be about 0.05\%. Using this isotopically engineered Si will
allow us to achieve only $N=5$ nuclear spins in the QD of typical size 
10$\times$10~nm. The mean distance between nuclear spins will be 
about 5~nm, which is
one order of magnitude larger than the lattice constant in Si. This large
distance prevents direct interaction between nuclear spins and make their
behavior independent.

We emphasize this fact because existence of interacting (by dipole
interactions) $N$ nuclear spins in a given dot leads to highly unwanted
multi-qubit gate. In this case, the conduction electron, which entangles two
nuclear spins each from a different dot, i.e. exciting one of $N$ nuclear
spins in a given dot, will produce of order of $2N$ excited states. In
contrast, non-interacting $N$ nuclear spins in a given dot form an analog of
a single nuclear spin qubit. The only problem is how to detect the quantum
computation process which consists in turnover of nuclear spin, if one
upturned nuclear spin coexists in a qubit with $N-1$ nuclear spins which
keep the previous spin values. For $N=5$, this means that the detected
signal will be only 20\% of the signal from the real single spin qubit
system. This is, in principle, solvable problem.

It is easy to show [13] that two nuclear spins, each from another dot,
entangled by a conduction electron, represents the logical
two-input-two-output SWAP gate, where outputs correspond to unchanged inputs
if both inputs are the same and turn over inputs if they are different:
\begin{center}
$|$00$\rangle $ $\rightarrow $ $|$00$\rangle $ $\ \ \ \ \ \ |$11$\rangle $ 
$\rightarrow $ $|$11$\rangle $

$|$01$\rangle $ $\rightarrow $ $|$10$\rangle $ \ \ \ \ \ \ $|$10$\rangle $ 
$\rightarrow $ $|$01$\rangle $
\end{center}

Let us assume that all $N$ nuclear spins in the 1st QD (dot A) are polarized
in ''down'' direction which corresponds to logical $|$0$\rangle ${}, while
in the second QD (dot B) all nuclear spins are polarized ''up'' which
corresponds to logical $|$1$\rangle ${}{}. So, the initial state of inputs
is $|$01$\rangle ${}{}. Since the Fermi (contact) hyperfine interaction is
``selective'', the electron-nuclear spin flip-flop process in a system of $N$
independent nuclear spins will take place between a given electron and a one
out of $N$ nuclear spins in the dot. Let us assume also that the conduction
electron has initially spin ''up''. The interaction between two nuclear
spins via the electron spin will occur as shown in fig.~4.

\begin{figure}[t]
\begin{center}
	\includegraphics{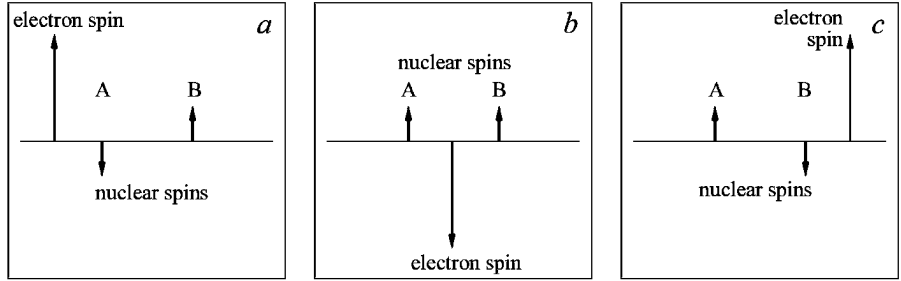}
	\caption{Two-qubit system as a SWAP gate
a) Initial state $|$01$\rangle ${}{}: Electron spin is ``up'', nuclei are: A
is ``down'' and B is ``up''; b) Intermediate state $|$11$\rangle $ after
electron-nuclear spin flip-flop interaction with dot A: electron spin is
``down'', nuclei are: A is ``up'' and B is ``up''; c) Final state $|$10$%
\rangle ${} after following electron-nuclear spin flip-flop interaction with
dot B: Electron spin is ``up'', nuclei are: A is ``up'' and B is ``down''.}
\end{center}
\end{figure}

Note, that because the electron Zeeman energy $\Delta E$ is more than three
orders of magnitude larger than the nuclear Zeeman splitting energy, the
intermediate state b) is energetically forbidden and is therefore virtual.
The Heizenberg uncertainty relation $\Delta E\Delta t$ $\sim h$ permits this
virtual state during the time interval $\Delta t$ $\sim h$/$\Delta E\sim $ 
10$^{-11}$~s. This allows the inter-dot distance up to 10$^2$--10$^3$~nm
which is much more than the maximal inter-dot limit in the proposed
technology. One can see also that virtual character of the intermediate
state makes impossible the change of nuclear spin direction in both dots if
they have the same nuclear spin polarization, never mind what is the
distance between dots.

\begin{figure}[t]
\begin{center}
	\includegraphics{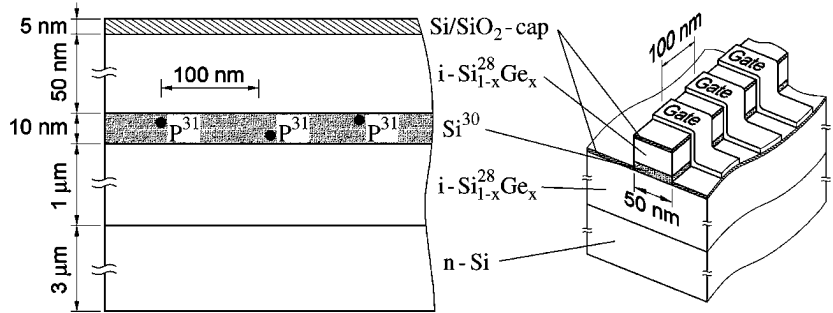}
	\caption{Architecture of the vertical
isotopically engineered Si/SiGe heterostructure with P atoms introduced
by NTD technique (left) and final Si-nanowire (right) prepared by
AFM-assisted lithography with array P atoms under metallic gates.}
\end{center}
\end{figure}

Thus, all conditions for the SWAP gate are fulfilled on the quantum level in
the proposed two-nuclear spin qubit system.

Enrichment of Si with isotope Si$^{30}$ followed by the
neutron-transmutation doping technique (NTD), allows also to suggest a
method for preparation of a prototype of an elementary nuclear-spin-qubit
embodied in a semiconductor based on fabrication of a isotopically
engineered Si$^{30}$/Si$^{28}$Ge heterostructure. In such a structure, P
atoms will be produced only within Si$^{30}$ strips, because Si$^{28}$
transmutes into stable isotope Si$^{29}$. Numerical estimations show that to
achieve the mean distance of $R\approx 100$~nm between P atoms in a 
Si$^{30}$ strip of 50$\times$10~nm cross-section, 
the integrated neutron flux must
be about 10$^{20}$ neutrons/cm$^2$ which corresponds to the relatively
long time (about one month) of irradiation in conventional nuclear reactors.
This is due to very small abundance of isotope Si$^{30}$ in the natural
composition (3.12\%). Enrichment of Si with isotope Si$^{30}$, for
example, up to 30\% will reduce the irradiation time down to a few days
which is quite reasonable. Large $R$ will allow us to easy arrange the
control gates. Entanglement between two separated P nuclear-spin-qubits in
such Si wire could be performed by mediation of the conducting electrons.
The mean distance $R$ between $P$ atoms will be controlled by the enrichment
of Si with isotope Si$^{30}$ and by integral dose of the neutron flux
irradiation. Spins associated with P donors will serve as qubits in
accordance with the Kane model [14,15]. Large distance $R\approx 100$~nm
will allow to easy arrange metallic gates above P atoms. Entanglement
between different qubits will be provided by mediation of conduction
electrons in a magnetic field. It is shown theoretically [16] that the
hyperfine interaction via the conducting electrons between nuclear spins
exhibit sharp maxima as a function of magnetic field and nuclear spin space
position. This phenomenon can be used for manipulation of qubits with almost
atomic precision.

The information encoded in the nuclear spin polarization could be detected
by an electrical measurement of two-electron system using
single-electron-transistor [17].  Fig.~5 shows the architecture of the 
Si/SiGe heterostructure prepared from different isotopes (Si layer is
prepared from Si$^{30}$, but two SiGe layers are prepared from isotope
Si$^{28}$) and the final Si-nano-wire with array of P atoms which is
expected to serve as qubits in S-MSQC.

\section*{Acknowledgments}

Author is thankful to I.D. Vagner, V.I. Safarov and P. Wyder for valuable
discussions. This work was financially supported by IST Programme of the
Commission of the European Communities, under contract number IST-2000-29686.


\newpage

\begin{thebibliography}{99}
\bibitem{1}  \textit{Neutron Transmutation Doping in Semiconductors}, ed. by
J. Meese (Plenum Press, NY, 1979).

\bibitem{2}  I.S. Shlimak, Physics of Solid State \textbf{41}, 716 (1999).

\bibitem{3}  H. Herrmann and H. Herzer, J. Electrochem. Soc. \textbf{122},
1568 (1975).

\bibitem{4}  \textit{Technology Roadmap for Nanoelectronics}, \ ed. by R.
Compa\~{n}\'{o}, \ L. Molenkamp, and D.J. Paul (European Comission, IST
programme, April 1999).

\bibitem{5}  W.S. Capinski, H.J. Maris, E. Bauser, I. Silier, M.
Asen-Palmer, T. Ruf, M. Cardona, and E. Gmenin, Appl. Phys. Lett. 
\textbf{71}, 2109 (1997).

\bibitem{6}  T. Ruf, R.W. Henn, M. Asen-Palmer, E. Gmelin, M. Cardona, H.J.
Pohl, G.G. Devyatykh, and P.G. Sennikov, Sol. State Commun., \textbf{115},
243 (2000).

\bibitem{7}  I. Shlimak, V.I. Safarov, and I.D. Vagner, J. Phys.: Condens.
Matter \textbf{13}, 6059 (2001).

\bibitem{8}  V. Privman, I.D. Vagner, and G. Kventsel, Phys. Lett. A 
\textbf{239}, 141 (1988).

\bibitem{9}  M. Doberts, K.v. Klitzing, J. Schneider, G. Weimann, and K.
Ploog, Phys. Rev. Lett. \textbf{61}, 1650 (1988).

\bibitem{10}  R.G. Mani, W.B. Jonson, V. Narayanamurti, V. Privman, and Y-H.
Zhang, Physica E \textbf{12}, 152 (2002).

\bibitem{11}  J.-L. Lin, D.Y. Petrovykh, J. Viernow, F.K. Men, D.J. Seo, and
F.J. Himsel, J. Appl. Phys. \textbf{84}, 255 (1998).

\bibitem{12}  F. Marchi, V. Bouchiat, H. Dallaporta, V. Safarov, D. Tonneau,
and P. Doppelt, J. Vac. Sci. Technol. B \textbf{16}, 2952 (1998).

\bibitem{13}  I. Shlimak and I.D. Vagner, in: \textit{Recent Trends in
Theory of Physical Phenomena in High Magnetic Fields}, 281 (Kluwer Academic
Publishers, 2003).

\bibitem{14}  B.E. Kane, Nature \textbf{393,} 133 (1998).

\bibitem{15}  B.E. Kane, Fortschr. Phys. \textbf{48}, 1023 (2000).

\bibitem{16}  Yu. Pershin, I.D. Vagner, and P. Wyder, J. Phys.: Condens.
Matter \textbf{15}, 997 (2003).

\bibitem{17}  B.E. Kane, N.S. Mac-Alpine, A.S. Dzurak, R.G. Clark, G.J.
Milburn, H.B. Sun,\ and H. Wiseman, Phys Rev B \textbf{61,} 2961 (2000).
\end{thebibliography}
\end{document}